\documentstyle[aps,prd,twocolumn]{revtex}

\begin{document}

\title{Gravitational collapse to toroidal, cylindrical and planar 
black holes 
\\
gr-qc/9709013
}

\author{Jos\'e P. S. Lemos}

\address{Departamento de Astrof\'{\i}sica,
	      Observat\' orio Nacional-CNPq, Rua General Jos\'e Cristino 77,
	      20921 Rio de Janeiro, Brazil \& \\
         Departamento de F\'{\i}sica,
	      Instituto Superior T\'ecnico, Av. Rovisco Pais 1, 
              1096 Lisboa, Portugal.}

\maketitle

\begin{abstract}
Gravitational collapse of non-spherical symmetric matter
leads inevitably to non-static external spacetimes. It is shown here
that gravitational collapse of matter with toroidal topology in a
toroidal anti-de Sitter background proceeds to form a toroidal black hole. 
According to the analytical model presented, 
the collapsing matter absorbs energy in the form of radiation 
(be it scalar, neutrinos, electromagnetic, or gravitational) from the exterior 
spacetime. Upon decompactification of one or two coordinates of the torus 
one gets collapsing solutions of cylindrical or planar matter onto black 
strings or black membranes, respectively. The results have implications on 
the hoop conjecture.  
\end{abstract}

\vskip 0.3cm

\noindent
{\bf 1. Introduction}

\vskip 1mm

Black hole (BH) solutions in an anti-de Sitter (adS) background whose 
event horizons have planar \cite{lemoscqg95}, cylindrical \cite{lemosplb95}, 
and toroidal topology \cite{lemosplb95,lemoszanchinprd96} have been 
found recently (see also \cite{huang95cai96brill97}).  
Since the importance of the spherical Schwarzschild BH 
has come from its role as the final state of complete gravitational 
collapse of a star, it is useful to investigate if these BHs with 
different topology may also emerge from gravitational collapse of some 
matter distribution. 

An important feature of spherical collapse onto a Schwarzschild BH is 
that, due to Birkhoff's theorem, spacetime is static outside the matter 
and the collapse proceeds without emission of gravitational waves. The 
same is true for the collapse of spherical matter in an adS background, 
i.e., in a spacetime with negative cosmological constant. On the other 
hand, it is well known that the collapse of cylindrical systems  
proceeds with emission of gravitational waves 
\cite{thorne72,piran78echeve93,apostolatosthorne93,nakamura97} which
creates additional problems in the modeling of these collapsing
systems.  It has also been known for a long time that collapsing
infinite dust cylinders form naked singularities \cite{thorne72}, not
BHs. This violates the cosmic censorship conjecture \cite{penrose69}
(which forbids the existence of singularities not surrounded by an
horizon), but not the hoop conjecture \cite{thorne72} (which states
that BHs form when and only when a mass $M$ gets compacted into a
region whose circumference in every direction is less than its
Scwharzschild circumference $4\pi M$ ($G=C=1$)). The cosmic censorship
is formulated to be applied to realistic gravitational collapse, which
in principle does not include cylindrical matter. However, in a certain
sense, cylindrical collapse can simulate the collapse of a finite
spindle \cite{shapiroteukolsky91}, near the central regions of the
spindle the collapse behaves as if the spindle is an infinite cylinder.
Besides these possible astrophysical applications, the collapse of
infinite cylinders probes and tests the structure of general
relativity.

In a previous work it has been conjectured by us 
\cite{lemoszanchinprd96} that, since there are known cylindrical BH 
solutions with a negative cosmological constant, collapse of 
cylindrical matter in a background with negative cosmological constant 
could form a cylindrical BH (i.e., a black string) rather than a naked 
singularity, violating in this way the hoop conjecture but not the 
cosmic censorship.  In this work we show that indeed cylindrical BHs 
(or black strings) form from gravitational collapse of cylindrical 
matter distribution. As the azimuthal cylindrical coordinate can be 
decompactified this solution also shows that planar BHs (or black 
membranes) can also form form gravitational collapse of a planar 
distribution. Moreover, since the `vertical' infinite cylindrical 
coordinate can, in turn, be compactified this solution also shows that 
toroidal BHs can form from gravitational collapse. Throughout this 
paper we will work mainly with the toroidal topology in mind, although 
the results can be modified straightforwardly to the other two cases, 
the main differences being the topologies themselves and the mass 
parameter which in the planar case is a surface mass density, in the  
cylindrical case is a linear mass density, and in the toroidal case it  
is a mass. In brief, we show that black membranes, black strings and 
toroidal BHs form from gravitational collapse.

Contrary to the spherical case, and as in the usual cylindrical 
collapse, the metric outside the collapsing toroidal (cylindrical 
or planar) matter is non-static. However, surprisingly, the problem 
can be solved exactly by using a modified Vaidya metric appropriate 
to the toroidal (cylindrical or planar) collapse (see section 2). 
Of course, this metric obeys Einstein field equations, and it 
describes the gravitational field associated with a toroidal 
(cylindrical or planar) flow of unpolarized scalar, neutrino, electromagnetic 
or gravitational radiation in the geometrical optics approximation. 
The interior solution we use is  a modified Friedmann 
solution also applicable to toroidal (cylindrical or planar) topology 
(section 3). By making a smooth matching at the interface (section 4) 
we find that the flux of waves modeled by the modified Vaidya metric is 
an incoming flux, and consequently that the mass parameter of the 
collapsing matter grows up to the formation of the BH.
By carefully choosing the right amout of incoming flux one avoids 
the emission of graviational waves from the collapsing matter. 
Finally, one can study the BH formation from the inside and outside 
points of view (section 5) and draw some conclusions (section 6).

\vskip .3cm

\noindent
{\bf 2. Exterior Radiating Solution}

\vskip 1mm

As stated in the introduction, 
the Schwarzschild metric represents the external field of a
collapsing spherical star, implying that in spherical collapse there is
no emission of gravitational waves.  Another spherically symmetric 
solution of Einstein field equations is the Vaidya metric which 
describes the gravitational field of an isotropic flow of unpolarized 
radiation in the geometrical optics approximation. It is usually 
employed in the study of imploding radiation shells \cite{joshi93}, as 
well as in modeling the external field of radiating stars 
\cite{misneretal65} and evaporating BHs \cite{york83}.  In the case of 
static or collapsing spherical stars, one can usually neglect the 
effects of this radiation and treat the external field as being given 
by the Schwarzschild metric. However, one should consider the Vaidya 
metric in the latest stages of the collapse when a supernova and a 
neutron star are formed accompanied by copious emission of neutrinos 
and photons.
It is also possible that a BH forms directly from the collapsing star 
without passing through the strong radiative stage, in which 
case the Schwarzschild metric gives again an accurate description for 
the external field. If the star is slightly non-spherical, the latest 
stages of such a direct collapse onto a BH produce some gravitational 
waves which have to be treated as a perturbation on the Schwarzschild 
spacetime. The effects of adding a negative cosmological constant do 
not alter radically the description. The main difference would be that 
the outside spacetime should be described by the Schwarzschild-adS 
metric or when relevant by the Vaidya-adS metric.

The situation changes drastically for a star with non-spherical topology. 
The reason being that, since there is no analogue of 
Birkhoff's theorem, a collapsing non-spherical symmetric star produces a 
non-static external spacetime. Notwithstanding, as we will show, one 
can treat the problem analitically for the gravitational collapse of 
toroidal (cylindrical or planar) configurations.

The Einstein field equations are 
\begin{equation}
G_{ab} + \Lambda g_{ab} = 8\pi  T_{ab}, 
                        \label{eq:2.1}
\end{equation}
where $G_{ab}$, $g_{ab}$, $T_{ab}$ are the Einstein, the metric and 
the the energy-momentum tensors, respectively, and $\Lambda$ is the 
cosmological constant ($G=C=1$). The equations admit the following solution 
\begin{equation}
ds^2 = - \left(\alpha^2r^2-\frac{q m(v)}{\alpha r}\right) dv^2 + 2 dv dr + 
r^2 (d\theta^2 + d\phi^2),
                         \label{eq:2.2}
\end{equation}
for an energy-momentum tensor given by
\begin{eqnarray}
&T_{ab} = \frac{q}{8\pi\alpha r^2} \frac{dm(v)}{dv} k_ak_b, &\nonumber \\
&  \quad k_a = - \delta^v_a, \quad k_ak^a = 0. &
                        \label{eq:2.3}
\end{eqnarray}
Here $\alpha \equiv \sqrt{\frac{-\Lambda}{3}}$, $v$ is the advanced 
time coordinate with $-\infty<v<\infty$, $r$ is the radial coordinate 
with $0<r<\infty$, and $\theta,\phi$ are the coordinates which describe 
the two-dimensional zero-curvature space generated by the 
two-dimensional commutative Lie group $G_2$ of isometries. The 
topologies of this two-dimensional space can be 
(i) $T^2=S^1 \times S^1$, the flat torus model  
$[G_2=U(1)\times U(1)]$, 
(ii) $R\times S^1$, the cylindricallly symmetric model 
$[G_2=R\times U(1)]$, and 
(iii) $R^2$, the planar model $[G_2=E_2]$. 
In the toroidal case we choose $0\leq\theta<2\pi$, $0\leq\phi<2\pi$, in 
the cylindrical case $-\infty<\theta<\infty$, $0\leq\phi<2\pi$, and in 
the planar case $-\infty<\theta<\infty$, $-\infty<\phi<\infty$. The 
parameter $q$ has different values depending on the topology of the 
two-dimensional space. For the torus $q=\frac{2\alpha}{\pi}$ and $m(v)$ 
is a mass, for the cylinder $q=4$ and $m(v)$ is a mass per unit length, 
and for the plane $q=\frac2\alpha$ and $m(v)$ is a mass per unit area.

Metric (\ref{eq:2.2}) is a modified Vaidya solution pertinent to toroidal 
(cylindrical or planar) topology. From the energy-momentum tensor given 
in equation (\ref{eq:2.3}) one can infer that it describes the gravitational 
field of a toroidal (cylindrical or planar) flow of unpolarized 
radiation in the geometrical optics approximation, and it can be 
used in modeling the external fields of matter with 
toroidal (cylindrical or planar) topology radiating or absorving
energy. Besides modeling radiation in the form of neutrinos or 
electromagnetic waves, $T_{ab}$ in (\ref{eq:2.3}) can also represent
scalar or gravitational radiation in an adequate limit. 
Now, noting that the energy-density of the radiation is 
$\epsilon = \frac{q}{8\pi\alpha r^2}\frac{dm}{dv}$, one sees that 
the weak energy condition for the radiation is satisfied whenever 
$\frac{dm}{dv}\geq0$, i.e., the radiation is imploding. 
For $m=$constant one has vacuum and equation (\ref{eq:2.2}) describes 
a static toroidal (cylindrical or planar) black hole  in ingoing (advanced 
time) Eddington-Finkelstein coordinates. The values of the parameter 
$q$ given above, were taken from ADM masses of the corresponding static 
BHs found in \cite{lemoscqg95,lemosplb95,lemoszanchinprd96}.

In these coordinates, lines with $v=$constant represent incoming radial 
null vectors whose generator vectors have the form $k^a=(0,-1,0,0)$, 
with $k_a = (-1,0,0,0)$ (see equation (\ref{eq:2.3})) . The generators 
$l^a$ of outgoing null lines do not have this simple form. By the 
conditions $l_al^a=0$ and $l_ak^a=-1$ one finds 
$l^a=(1,\frac12(\alpha^2r^2-\frac{qm(v)}{\alpha r}),0,0)$. 
The equation for outgoing radial null geodesics $r(v)$ can easily be find 
by puting $ds^2=0$ in metric  (\ref{eq:2.2}), yielding, 
\begin{equation}
\frac{dr}{dv} = \frac12 (\alpha^2 r^2 - \frac{qm(v)}{\alpha r}).  
                        \label{eq:2.4}
\end{equation}
The exterior solution discussed here is to be matched to the 
interior metric presented in the following section. 

\vskip .3cm

\noindent
{\bf 3. Interior Matter Solution}
\vskip 1mm
We now assume that the interior spacetime is made of a homogeneous 
collapsing dust cloud whose spacetime is described through 
a flat Friedmann-Robertson-Walker type metric given by 
\begin{equation}
ds^2 = -dt^2 + a(t)^2 \left(dl^2 + l^2(d\theta^2+d\phi^2)\right), 
                        \label{eq:3.1}
\end{equation}
where $t$ and $l$ are comoving coordinates, and again, $\theta,\phi$ 
are the coordinates which describe the two-dimensional zero-curvature 
space (torus, cylinder or plane). Due to Einstein field equations  
(\ref{eq:2.1}) we have to set to zero the usual Friedmann-Robertson-Walker 
curvature parameter, $k=0$ without loss of generality. For dust the 
energy-momentum tensor is given by 
\begin{equation}
T_{ab} = \rho u_au_b,  
                        \label{eq:3.2}
\end{equation}
where $\rho$ is the energy-density of the matter, and $u^a$ its 
four-velocity. Einstein field equations (\ref{eq:2.1}) yield
\begin{equation}
\frac{1}{a^2}\left(\frac{da}{dt}\right)^2 = \frac{8\pi}{3} \rho - \alpha^2 , 
                        \label{eq:3.3}
\end{equation}
\begin{equation}
\frac1a \frac{d^2a}{dt^2} = - \frac{1}{2a^2}\left(\frac{da}{dt}\right)^2 
- \frac{3\alpha^2}{2}, 
                        \label{eq:3.4}
\end{equation}
where again $\alpha \equiv \sqrt{\frac{-\Lambda}{3}}$. 
Integration of these equations give 
\begin{equation}
\rho =  \frac{\rho_0}{\sin^{2}\left( \frac32 \alpha t \right)},
                        \label{eq:3.5}
\end{equation}
and
\begin{equation}
a = a_0 \sin^{2/3}\left( \frac32 \alpha t \right), 
                        \label{eq:3.6}
\end{equation}
where $\rho_0$ and $a_0$ are constants, with $\rho_0 = 
\frac{3\alpha^2}{8\pi}$. Restoring the constant $G$ one has  $\rho_0 = 
\frac{3\alpha^2}{8\pi G}$, which means the initial density is 
independent of the mass and radius of the initial configuration (a 
similar situation was found in \cite{ilhalemos97} for gravitational
collapse in Lovelock gravity). This could be expected, since for very  
small $\alpha$ one finds $\rho = \frac{1}{6\pi G t^2}$, recovering the 
flat model with toroidal topology and (almost) zero cosmological 
constant, similar in its time-dependence to the spherical 
flat-Friedmann model.  Note also that $\rho a^3$= constant.  
Defining $\overline{t} \equiv \frac32 \alpha t$, one sees that the
solution is valid in the time range $0<\overline{t}<\pi$. From
$0<\overline{t}<\pi/2$ the matter is expanding, from
$\pi/2<\overline{t}<\pi$ the matter is collapsing.  At
$\overline{t}=\pi/2$ there is a moment of time symmetry. We are
interested in the collapsing part of the solution, and thus we take
$\pi/2\leq \overline{t}<\pi$. The energy-density of the matter as well
as the Kretschmann scalar blow up at $\overline{t}=\pi$, indicating
the formation of a spacetime singularity.

\vskip .3cm

\noindent
{\bf 4. Matching}
\vskip 1mm

To match the interior and exterior spacetimes, across an interface 
of separation $\Sigma$, we use the junction conditions 
\begin{eqnarray}
\left. ds^{2}_{+}\right]_{\Sigma} = \left. ds^{2}_{-}\right]_{\Sigma} \label{eq:4.1} \\ 
\left. K_{ab}^{+}\right]_{\Sigma} = \left. 
K_{ab}^{-}\right]_{\Sigma} \label{eq:4.2}
\end{eqnarray}
where $K_{ab}$ is the extrinsic curvature, 
\begin{equation}
K_{ab}^{\pm}= -n_{e}^{\pm}\,\frac{\partial^{2}
x_{\pm}^{e}}
{\partial \xi^{a}\partial \xi^{b}} - n_{e}^{\pm}\,
\Gamma_{cd}^{e}\,
\frac{\partial x_{\pm}^{c}}{\partial \xi^{a}}\,
\frac{\partial x_{\pm}^{d}}{\partial \xi^{b}}
\label{eq:4.3}
\end{equation}
and $n_{e}^{\pm}$ are the components of the unit normal vector to
$\Sigma$ in the coordinates $x_{\pm}$, and $\xi$ represents 
the intrinsic coordinates in $\Sigma$.  
The subscripts $\pm$
represent the quantities taken in the exterior and interior
spacetimes.  Both the metrics and the extrinsic curvatures in
(\ref{eq:4.1})-(\ref{eq:4.2}) are evaluated at $\Sigma$.  It is 
useful to define a metric
intrinsic to $\Sigma$  as 
\begin{equation}
ds^2_{\Sigma} = -d\tau^2 + R^2(\tau)\,
(d\theta^2 + d\phi^2)  ,
\label{eq:4.4}
\end{equation}
where $\tau$ is the proper time on $\Sigma$.

We analyse first the surface $\Sigma$ as viewed from the exterior 
spacetime. To match the exterior metric with the metric on 
$\Sigma$ we use the junction condition (\ref{eq:4.1}) 
and metrics  (\ref{eq:2.2}),  (\ref{eq:4.4}), to obtain 
\begin{equation}
\left. r(v)\right]_{\Sigma} = R\left(\tau \right) \, ,
\label{eq:4.5a}
\end{equation}
and 
\begin{equation}
\left. \alpha^2r^2 - \frac{qm(v)}{\alpha r} - 2 \frac{dr}{dv} 
\right]_{\Sigma} = 
\left. \frac{1}{\left(dv/d\tau\right)^2} \right]_{\Sigma}, 
                        \label{eq:4.5b}
\end{equation}
where both equations are evaluated on $\Sigma$.  
The unit normal to $\Sigma$ in the exterior spacetime is 
\begin{equation}
n_e^+ = \frac{1}{\sqrt{-2\frac{dr_\Sigma}{dv} + \alpha^2r_\Sigma^2 - 
\frac{qm(v)}{\alpha r_\Sigma}}}
\left(-\frac{dr_\Sigma}{dv},1,0,0\right). 
                        \label{eq:4.6a}
\end{equation}
Using  (\ref{eq:4.5b}) and $\frac{dr}{dv}=\frac{dr/d\tau}{dv/d\tau}$ 
we can put (\ref{eq:4.6a}) in the form 
\begin{equation}
n_e^+ = \left(-\frac{dr_\Sigma}{d\tau},\frac{dv_\Sigma}{d\tau},0,0\right). 
                        \label{eq:4.6b}
\end{equation}
From now on, we will usually omit the subscript $\Sigma$ to 
denote evaluation at the interface.
From (\ref{eq:4.3}) we then get the $K_{ab}^+$ component of interest: 
\begin{equation}
K_{\theta\theta}^{+} = -r \frac{dr}{d\tau} + r \frac{dv}{d\tau} 
\left(\alpha^2 r^2 - \frac{qm(v)}{\alpha r}\right), 
                        \label{eq:4.7}
\end{equation}
valid on $\Sigma$, of course. 

We now analyse the surface $\Sigma$ as viewed from the interior spacetime.  
To match the interior metric with the metric on 
$\Sigma$ we use the junction condition (\ref{eq:4.1}) 
and metrics  (\ref{eq:3.1}),  (\ref{eq:4.4}), to obtain
\begin{equation}
l_\Sigma a(t) = R(\tau) 
                        \label{eq:4.8a}
\end{equation}
and 
\begin{equation}
\frac{d\tau}{dt} = 1,  
                        \label{eq:4.8b}
\end{equation}
valid on $\Sigma$. 
The unit normal to $\Sigma$ in the interior spacetime is
\begin{equation}
n_{e}^{-} = \left(0, a(t), 0, 0\right). 
                        \label{eq:4.9}
\end{equation}
From  (\ref{eq:4.3}) we then have
\begin{equation}
K_{\theta\theta}^{-} = l_\Sigma a(t) = R(\tau). 
                        \label{eq:4.10}
\end{equation}

In order to have a smooth matching one imposes that the extrinsic curvatures 
(\ref{eq:4.7}) and (\ref{eq:4.10}) must be equal, 
$K_{\theta\theta}^{+} = K_{\theta\theta}^{-}$, yielding 
\begin{equation}
-r \frac{dr}{d\tau} + r \frac{dv}{d\tau} 
\left(\alpha^2 r^2 - \frac{qm(v)}{\alpha r}\right) = R(\tau) , 
                        \label{eq:4.11}
\end{equation}
an equation valid on $\Sigma$. Using  (\ref{eq:4.11}) and (\ref{eq:4.5b})
we find 
\begin{equation}
\frac{dv}{d\tau} = \frac{1}{1-dR/d\tau}, 
                        \label{eq:4.12}
\end{equation}
on $\Sigma$. Using now (\ref{eq:4.11}), (\ref{eq:4.5b}), (\ref{eq:4.12}) 
and (\ref{eq:3.3}) we have 
\begin{equation}
m(v) = \frac{\alpha}{q} \left( \frac{8\pi}{3}\rho R^3 - 
R(\tau)\right). 
                        \label{eq:4.13}
\end{equation}
Recalling from equations (\ref{eq:3.5})-(\ref{eq:3.6}) that 
$\rho R^3=\rho_0 R_0^3$ with $R=a(t)l_\Sigma$ (see equation 
(\ref{eq:4.8a})), and using (\ref{eq:4.5a}) we find from 
(\ref{eq:4.13})
\begin{equation}
m(v) = \frac{\alpha}{q} \left( \frac{8\pi}{3}\rho_0 R_0^3 - 
r_\Sigma(v)\right), 
                        \label{eq:4.14}
\end{equation}
which gives the evolution of the mass with the external time $v$. Equation 
(\ref{eq:4.14}) is the most important result of the section. 
By (\ref{eq:4.8a}), we know $R(\tau) = l_\Sigma a(t)$, and from 
(\ref{eq:3.6}) we have the evolution of the radius of the surface of the 
star with time $t$ (or $\tau$, since $dt=d\tau$). 
Then from (\ref{eq:4.12}) one gets $v = v(\tau)$, or its inverse 
$\tau = \tau(v)$.  
Since from equation (\ref{eq:4.5a}) $r_\Sigma(v)=R(\tau)$, we find that 
$r_\Sigma (v)$ is a known function of $v$. Unfortunately, integration 
of  (\ref{eq:4.12}) 
to find $v(\tau)$ cannot be performed analytically. However, since for 
collapse $\frac{dR}{d\tau}<0$ we find that $v$ increases monotonically 
with $\tau$ or $t$. Thus, as the collapse proceeds $r_\Sigma$ decreases 
from its maximum value ${r_\Sigma}_0\equiv r_\Sigma(v_0)=R_0$, 
where $v_0$ denotes the time at the onset of the collapse. 
Therefore, we have obtained the result that the mass of the 
cloud $m(v)$ increases during the collapse due to the incoming 
flux of the high frequency radiation, be it in the form of scalar, 
neutrinos, electromagnetic or gravitational waves. 

It is convenient to define an initial mass $m_0$ in terms of $\rho_0$ 
and ${r_\Sigma}_0 = R_0$ through the equation
\begin{equation}
\frac{qm_0}{\alpha} = \frac{8\pi}{3} \rho_0 {r_\Sigma}_0^3 - {r_\Sigma}_0. 
                        \label{eq:4.15}
\end{equation}
Then equation (\ref{eq:4.14}) can be written as 
\begin{equation}
m(v) = m_0 + \frac{\alpha}{q}\left( {r_\Sigma}_0 - r_\Sigma(v) \right). 
                        \label{eq:4.16}
\end{equation}
This equation will be used in the next section.

\vskip .3cm

\noindent
{\bf 5. Black Hole Formation}
\vskip 1mm

To study BH formation we distinguish two situations, the inside and 
outside stories. 

For the interior of the star we use metric (\ref{eq:3.1}) and study
collapse in the range  $\pi/2\leq \overline{t}<\pi$ 
($\overline{t}\equiv\frac32 \alpha t$). At the onset of
the collapse, at time $\overline{t}=\pi/2$, there are no singularities. The
singularity forms at $\overline{t}=\pi$ where the curvature scalars and the
density blow up (see equation  (\ref{eq:3.6})). The appearance of an
apparent horizon indicates the formation of a BH. Here, the apparent
horizon is defined as the boundary of trapped two-tori 
(two-cylinders or two-planes) in spacetime. To
find this boundary we look for two-tori (two-cylinders or two-planes)
$Y\equiv a(t) l=$constant whose
outward normals are null, i.e., $\nabla Y \cdot \nabla Y = 0$, yielding
$l_{\rm AH}=-\frac{1}{da/dt}$. Using equation  (\ref{eq:3.6}) this is,  
\begin{equation}
\frac{l_{\rm AH}}{l_\Sigma} = - 
\left(\frac{3}{8\pi\rho_0R_0^3 \alpha}\right)^{1/3} 
\frac{\sin^{1/3}\frac32 \alpha t}{\cos \frac32 \alpha t}.
                        \label{eq:5.1}
\end{equation}
The apparent horizon first forms at the surface of the star 
$l_{\rm AH} = l_\Sigma$. Thus, for given $\rho_0, R_0$ and $\alpha$ 
one can find from equation (\ref{eq:5.1}) the time $t$ at which the apparent 
horizon first forms.

For outside observers the descripition is different as they should use 
metric (\ref{eq:2.2}). If one looks for trapped two-tori 
(two-cylinders or two-planes) whose outwards 
normals are null, $\nabla r \cdot \nabla r =0$, one now obtains the 
condition
\begin{equation}
\alpha^3 r_{\rm AH}^3 = q m(v). 
                        \label{eq:5.2}
\end{equation}
Differentiating equation  (\ref{eq:5.2}) and puting it back in the metric 
(\ref{eq:2.2}) gives 
$
ds^2=\frac{2q}{3\alpha^3r^2}\frac{dm(v)}{dv} dv^2 + r^2 (d\theta^2 + d\phi^2). 
$ 
The sign of $dm(v)/dv$ decides on the character of the AH. 
Here $dm(v)/dv>0$. Thus, the apparent horizon is an unphysical 
spacelike surface, interior to the the surface of the matter 
where the metric (\ref{eq:2.2}) is not valid. Equation (\ref{eq:5.2}) 
is only valid at $v=v_{\rm AH}$, the time at which the apparent 
horizon forms.

We have a dynamic situation. As the matter collapses, 
the mass of the toroidal star increases. To find the radius 
at which the apparent horizon forms we equate equation (\ref{eq:5.2}) 
to (\ref{eq:4.16}) to obtain a cubic equation for $r_{\rm AH}$ 
\begin{equation}
\alpha^3 r_{\rm AH}^3 + \alpha r_{\rm AH} - 
\left( qm_0 + \alpha {r_\Sigma}_0 \right) = 0. 
                        \label{eq:5.3}
\end{equation}
This equation has one real root. Defining $C$, $A$ and $B$ as 
\begin{equation}
C=qm_0 + \alpha {r_\Sigma}_0,
                        \label{eq:5.4}
\end{equation}
\begin{equation}
A= ^{1/3}\sqrt{ \frac{C}{2} + \sqrt{ \frac{C^2}{4} + \frac{1}{27} }   },
                        \label{eq:5.5}
\end{equation}
and 
\begin{equation}
B = ^{1/3}\sqrt{ \frac{C}{2} - \sqrt{ \frac{C^2}{4} + \frac{1}{27} }   },
                        \label{eq:5.6}
\end{equation}
the solution for $r_{\rm AH}$ can be written as
\begin{equation}
\alpha r_{\rm AH} = A + B. 
                        \label{eq:5.7}
\end{equation}

One can check that both procedures (the inside and outside stories) 
give the same time $t$ (and $v$) 
for the formation of the apparent horizon. As an example we 
choose ${r_\Sigma}_0 = \frac12$, $\rho_0 = \frac5\pi$, $\alpha=\frac23$, 
$qm_0=\frac79$. Then puting $l_{\rm AH} = l_\Sigma$ in (\ref{eq:5.1}) 
gives $t=2.51$. On the other hand, one gets from (\ref{eq:5.7}) 
$r_{\rm AH}=1.09$, which upon using (\ref{eq:3.6}) suitably gives 
again $t=2.51$.

There is an interesting situation in toroidal collapse. The initial 
configuration can collapse  from a negative mass parameter $m_0$ into a 
BH with positive mass.  To find the minimum possible mass that yields a 
BH we set 
\begin{equation}
m(v_{\rm AH}) \equiv m_{\rm AH} =0 . 
                        \label{eq:5.8}
\end{equation}
Puting this condition back in equation (\ref{eq:4.16}) implies 
\begin{equation}
r_{\rm AH} \equiv r(v_{\rm AH}) = {r_\Sigma}_0 + \frac{q}{\alpha} m_0. 
                        \label{eq:5.9}
\end{equation} 
Equations  (\ref{eq:5.9}) and  (\ref{eq:5.3}) then give $r_{\rm AH}=0$.
Thus, $m_{\rm AH}=0$ implies $r_{\rm AH}=0$. In addition, equation 
(\ref{eq:5.9}) gives that the initial mass should obey 
\begin{equation}
m_0 \geq - \frac{\alpha}{q} {r_\Sigma}_0. 
                        \label{eq:5.10}
\end{equation} 
Below this value of $m_0$ the collapsing matter forms a naked singularity 
rather than a BH.

To deal with the cubic equation (\ref{eq:5.3}) one can study two limiting 
situations. First, assume $C = q m_0 + \alpha {r_\Sigma}_0 << 1$ 
with $\alpha {r_\Sigma}_0 >> q m_0$. Then the solution is
\begin{equation}
\alpha r_{\rm AH} = \alpha {r_\Sigma}_0 \left( 1 - \alpha^2 
{r_\Sigma}_0^2 + 
\frac{qm_0}{\alpha {r_\Sigma}_0}\right).
                        \label{eq:5.11}
\end{equation}
In order that there is no black hole at the onset of the collapse 
one has $\alpha^2 {r_\Sigma}_0^2 > \frac{qm_0}{\alpha {r_{\Sigma}}_0}$, 
which implies ${r}_{\rm AH} < {r_\Sigma}_0$ always. 

Second,  if we assume $C = q m_0 + \alpha {r_\Sigma}_0 >> 1$ with 
$\alpha^3 {r_\Sigma}_0>C$ then the solution for the apparent horizon is 
\begin{equation}
\alpha r_{\rm AH} = \left(qm_0 + \alpha {r_\Sigma}_0\right)^{1/3} - 
\frac{1}{3\left(qm_0 + \alpha {r_\Sigma}_0 \right)^{1/3}}, 
                        \label{eq:5.12}
\end{equation}
where again ${r}_{\rm AH} <  {r_\Sigma}_0$, as it should.

We still have to check that the collapsing star fades from sight 
to external observers. A null geodesic emerging form the surface 
of the star $r_\Sigma$ at a time $v$ arrives at an observer at 
point $r$ at time $v_{\rm obs}$ given by 
\begin{equation}
v_{\rm obs} = v + {\int}_{r_\Sigma}^r \frac{dv}{dr} dr. 
                        \label{eq:5.13}
\end{equation}
Using equation  (\ref{eq:2.4}) this gives 
\begin{eqnarray}
&v_{\rm obs} = v + \frac{2}{\alpha} \frac{1}{(qm(v))^{1/3}} 
\left[ 
\frac16 \ln\left(\alpha r - (qm(v))^{1/3}\right)^2 - \right. &\nonumber \\
&
\frac16 \ln\left(\alpha^2 r^2 + \alpha r (qm(v))^{1/3}+ (qm(v))^{2/3}\right) \\ 
&\left.
+\frac{1}{\sqrt3} \tan^{-1} \frac{2\alpha r + 
(qm(v))^{1/3}}{\sqrt3 (qm(v))^{1/3}}
\right]^r_{r_\Sigma}. & 
                        \label{eq:5.14}
\end{eqnarray}
Then $v_{\rm obs}\rightarrow \infty$    as the surface of the 
star approaches the apparent horizon, 
$r_\Sigma\rightarrow \frac{qm(v_{\rm AH})}{\alpha}$. As seen by outside 
observers the collapse to a BH takes an infinite time. One can also 
check that the redshift $z$ of the light diverges as the apparent horizon 
is approached,  $z\propto -\frac{dr_\Sigma/dt}{r_\Sigma - (qm)^{1/3}}$, 
implying that light from star gets ever dimmed. The solution (\ref{eq:2.2}) 
will then tend to the static BH solutions found in 
\cite{lemoscqg95,lemosplb95,lemoszanchinprd96}.

\vskip .3cm

\noindent
{\bf 6. Conclusions}

\vskip 1mm
We have found collapsing solutions of toroidal (cylindrical, or planar) 
matter onto toroidal (cylindrical, or planar) 
BHs. In the model used there is a flux of high frequecy radiation 
towards the star. This incoming radiation has to be chosen with the right 
flux (see equation (\ref{eq:4.14})) in order to avoid the emission of 
gravitational waves. If there is no such precisely chosen incoming flux, 
the toroidal (cylindrical or planar) configuration will presumably collapse 
to form a BH with emission of large amounts of gravitational waves 
\cite{ref97}.

For external observers, the solution tends asymptotically
with time to static BH solutions.
The maximally analytical extension of these static BH solutions show
unphysical regions, such as the white hole region behind the past event
horizon, which also appear in the Schwarzschild solution.
When one performs complete gravitational collapse of some matter 
configuration in an approprite background, this part
of the solution will cover the unphysical regions of the maximally
extended black hole. In the corresponding Penrose diagram, each point 
representing a torus, a cylinder or a plane, depending on the topology 
chosen, one visualizes collapsing solutions of toroidal, cylindrical or planar 
matter onto black holes, black strings or black membranes, respectively. 
Collapse to BH solutions with pseudospherical horizons have also been 
recently studied \cite{aminneborg96mann97}. 

A further extension of this paper could be the inclusion of the 
Teichmuller complex parameter, which specifies conformally equivalent 
classes of the torus, into the metric of the toroidal BH \cite{vanzo} 
and see if it produces significant changes.

The Vaidya metric is also used to study the formation of naked 
singularities in spherical gravitational collapse \cite{joshi93}. 
In a preliminary study \cite{lemos97} we have found that collapse of toroidal 
(cylindrical or planar) radiation using the modified Vaidya metric 
of section 2 does not yield naked singularities, strengthening the claim 
made in the introduction that non-spherical collapse in a 
negative cosmological constant background may violate the hoop 
but not the cosmic censorship conjecture.

\end{document}